\documentclass{aipproc}

\layoutstyle{8x11single}

\begin{document}

\title{Results in Optimal Discrimination}
\author{Kieran Hunter}{address={University of Strathclyde, Glasgow, UK}}

\begin{abstract}
We study the problem of discriminating between non-orthogonal quantum states with least probability of error. We demonstrate that this problem can be simplified if we solve for the error itself rather than solving directly for the optimal measurement. This method enables us to derive solutions directly and thus make definite statements about the uniqueness of an optimal strategy. This approach immediately leads us to a state-discrimination analogue of Davies Theorem \cite{Davies}.

In the course of this, a complete solution for distinguishing equally likely pure qubit states is presented.
\end{abstract}
\maketitle

One cannot perfectly discriminate between non-orthogonal quantum states. It is therefore sensible to ask how one can {\em best} discriminate such states. 

We consider the situation where we know that a system is in one of a pre-determined set of states $\{\hat{\rho}_{j}\}$, each of which occurs with probability $p_{j}$, but we do not know which one. Such a situation might arise if we knew the properties of the preparation device (or communications protocol), but not the actual setting used on this specific preparation. We can make any possible measurement to identify the state, but we {\em must} make an identification of the state based on the result of the measurement. If the states $\{\hat{\rho}_{j}\}$ are not mutually orthogonal, then there will be a non-zero probability that this identification will be wrong. This is the state discrimination problem. 

In this situation it makes sense to define the optimal measurement strategy to be the one which minimises the probability $P_{e}$ of incorrectly identifying the state. This is the most well known example of a hypothesis testing problem, as detailed in \cite{Helstrom, Holevo, Yuen}.

We describe our measurement strategy with a Probability Operator Measure (POM). This is a set of operators 
$\{\hat{\Pi}_{k}\}$ which gives the probabilities of each possible measurement outcome $P(k|j)$:
\begin{equation}
P(k|j) = \mathrm{Tr}\left(\hat{\Pi}_{k}\hat{\rho}_{j}\right)
\end{equation}
when the system is in the state $\hat{\rho}_{j}$.
The elements $(\hat{\Pi}_{k})$ of the POM represent probabilities and therefore must be subject to the following conditions:
\begin{equation}
\mathrm{Non-negativity: } \ \hat{\Pi}_{k} = \hat{\Pi}_{k}^{+} \geq 0 \  \forall \ k \label{pos}
\end{equation}
and
\begin{equation}
\mathrm{Completeness: } \sum_{k} \hat{\Pi}_{k} = \hat{1}. \label{comp}
\end{equation}
Each POM element $(\hat{\Pi}_{k})$ corresponds to the detection of the state $\hat{\rho}_{k}$. Thus there must be exactly as many POM elements as there are possible states, though some of these POM elements may be zero operators corresponding to states which are never detected.

\section{Minimising the Error Probability}

For a measurement strategy to minimise the error probability $P_{e}$, it must satisfy a known set of necessary and sufficient conditions 
\cite{Helstrom, Holevo, Yuen}:
\begin{equation}
\sum_{k} p_{k}\hat{\rho}_{k}\hat{\Pi}_{k} -  p_{j}\hat{\rho}_{j} \geq 0 \ \forall \ j. \label{nscon}
\end{equation}
It is possible to derive \cite{Helstrom, Yuen} a useful necessary condition on the minimum error strategy:
\begin{equation}
\left(\hat{C} -  p_{k}\hat{\rho}_{k}\right)\hat{\Pi}_{k} = 0 \ \forall \ k \label{nec}
\end{equation}
where 
\begin{equation}
\hat{C} =  \sum_{j} p_{j}\hat{\rho}_{j}\hat{\Pi}_{j}, \label{Cdef}
\end{equation}
from the necessary and sufficient conditions (\ref{nscon}) and the completeness condition (\ref{comp}). This necessary condition is equivalent to stating that the first derivative of the error probability is zero \cite{Helstrom, Holevo}.

While these conditions do give us a starting point for finding minimum error POMs, they do not themselves provide a great insight into either the form of minimum error measurement strategies, or into how error probability depends on the set of possible states. For this we must examine the solutions to these conditions, and there are not many solutions.

The solved cases of the necessary and sufficient conditions can be categorised:
\begin{enumerate}
\item When there are only two possible states \cite{Helstrom}
\item When the states possess some simplifying symmetry \cite{Yuen, Symerr, Eldar, Barsym, Erika1}
\item When we have states for which the error cannot be reduced by measurement \cite{Nome}.
\end{enumerate}
Only the first of these solutions is directly derived from the conditions (\ref{nscon}), the others were all postulated and then shown to satisfy these conditions. Simply postulating a solution does not tell us whether that solution is unique. This lack of uniqueness does not aid our goal of understanding the irreducible error of measurements on non-orthogonal quantum states.

%\section{Solving the Minimum Error Conditions}

The essential difficulty in solving the conditions directly (rather than postulating solutions) is that all of the variables $(\hat{\Pi}_{k})$ appear in each condition, and they are not independent variables.\footnote{This is also why the two states case is more easily soluble: there is only one independent variable.} 
However, the operator $\hat{C}$ (in (\ref{nec}) and (\ref{Cdef})) fixes the error probability $P_{e}$ as $P_{e} = 1-\mathrm{Tr}(\hat{C})$. This is a clue as to how we might solve the necessary and sufficient conditions: we will solve them for $\hat{C}$ rather than for $\{\hat{\Pi}_{k}\}$.\footnote{This is rather reminiscent of the method used in \cite{Yuen} to prove the necessity of the conditions (\ref{nscon}).}

We can see that the necessary condition (\ref{nec}), combined with the POM conditions (\ref{comp}), implies the equality (\ref{Cdef}) originally used as the definition of $\hat{C}$. Thus we can use these conditions as an alternative definition of $\hat{C}$. Furthermore, since $\mathrm{Tr}(\hat{C})$ fixes the error probability, all strategies $\{\hat{\Pi}_{k}\}$ which satisfy the necessary condition (\ref{nec}) for the optimal $\hat{C}$ will be optimal strategies. 

We can now restate the problem as finding an operator $\hat{C}$ such that:
\begin{equation}
\left(\hat{C} -  p_{k}\hat{\rho}_{k}\right)\hat{\Pi}_{k} = 0 \ \forall \ k, \label{Cnec}
\end{equation}
and, from (\ref{nscon}),
\begin{equation}
\hat{C} -  p_{k}\hat{\rho}_{k} \geq  0 \ \forall \ k, \label{Cns}
\end{equation}
for some POM $\{\hat{\Pi}_{k}\}$. Here $\hat{C}$ is our operator variable, and the optimal POM is derived from $\hat{C}$ by using (\ref{Cnec}), (\ref{pos}) and (\ref{comp}).

From this we can immediately see that the optimal strategy might not be unique. The relation between the error 
$(\hat{C})$ and the measurement $\{\hat{\Pi}_{k}\}$ is fixed by (\ref{Cnec}), but (\ref{Cnec}) places no restriction on 
$\{\mathrm{Tr}(\hat{\Pi}_{k})\}$. The only restriction on these traces are the POM conditions (\ref{pos}) and 
(\ref{comp}), and these conditions cannot uniquely determine all of the $\{\mathrm{Tr}(\hat{\Pi}_{k})\}$ when the number of POM elements is large.

For the purpose of actually finding the solutions, this formulation suggests a two-step procedure:
\begin{enumerate}
\item Find operators $\hat{C}$ which can satisfy (\ref{Cnec}) and (\ref{Cns}), then
\item Check which of these operators leads to elements which can form a POM.
\end{enumerate}
An important restriction on $\hat{C}$ for step (1) is that, from (\ref{Cnec}), we must have either
\begin{equation}
\mathrm{Det}\left(\hat{C} -  p_{k}\hat{\rho}_{k}\right) = 0  \ \mathrm{or} \ \hat{\Pi}_{k} = 0 \  \forall \ k. \label{det}
\end{equation}
%
%for each state $\hat{\rho}_{k}$. 
Thus, for each outcome which occurs with non-zero probability, the corresponding $\hat{C} -  p_{k}\hat{\rho}_{k}$ must have a zero eigenvalue.
For discriminating between sets of qubit states this approach becomes especially useful, since then $\hat{C} -  p_{k}\hat{\rho}_{k}$ and $\hat{\Pi}_{k}$ must be proportional to pure state projectors. 

We will demonstrate a simple example of how one use this method to find not {\bf an} optimal strategy, but {\bf \em all} optimal strategies for a given class of states.

\section{Example: Distinguishing Equiprobable Pure Qubit States}

Let us consider the problem of discriminating between a set of $N$ equally likely pure qubit states. We seek to find all solutions to the necessary and sufficient conditions (\ref{nscon}) for such sets of states. We use the following method: 
\begin{description}
\item We assume that we do not {\em need} zero operators as POM elements (see (\ref{det})),
\item We solve (\ref{det}) for $\hat{C}$,
\item We obtain the directions of the POM elements from (\ref{Cnec}),
\item We solve (\ref{comp}) for $\mathrm{Tr}(\hat{\Pi}_{k})$.
\end{description}
If no solution is found, we know that at least one POM element must be a zero operator. We then update our assumption accordingly, and try again. We shall begin by looking at a set of three states.

\begin{figure}[ht]
\resizebox{0.4\linewidth}{!}{\includegraphics*{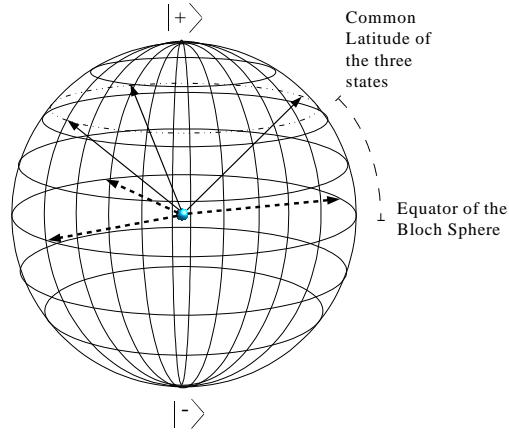}}
\caption{The position on the Bloch sphere of the projectors proportional to the optimal POM elements (dashed arrows) for
discriminating between a set of three equiprobable pure qubit states (solid arrows). These elements are
equatorial in the basis where the states share a common latitude and each element has the
same longitude as the corresponding state. These elements form a POM if and only if every
semicircle of the circle of common latitude contains at least one state.} \label{fig}
\end{figure}

The most useful basis to describe the three states in is the basis where the diagonal elements of their density operators are identical. In this basis the three states share a common latitude of the Bloch Sphere (see figure \ref{fig}). If we then follow the procedure outlined above, we find that (\ref{det}) is insufficient to completely define the optimal $\hat{C}$, but the diagonal part of (\ref{comp}) fixes the rest.

The solution for three states (see figure \ref{fig}) is similar to that for the symmetric states \cite{Symerr} in that the optimal POM elements are at the Bloch Sphere longitudes of the states and equatorial in latitude, providing that such elements can satisfy 
(\ref{comp}) and so form a POM. $\hat{C}$ is diagonal in the basis of figure \ref{fig}, but it does not correspond to the square-root strategy \cite{Symerr, Eldar, Barsym} unless the three states actually {\em are} symmetric states as defined by \cite{Symerr}.

It is interesting to note that the minimum error in this case depends only on the common latitude of the three states and not on their arrangement on that latitude:
\begin{equation}
\mathrm{min}(P_{e}) = 1-p_{j}-2p_{j}\sqrt{\langle +|\hat{\rho}_{j}|+ \rangle\langle -|\hat{\rho}_{j}|- \rangle},
\end{equation}
which is the same for any of the $\hat{\rho}_{j}$s. It clearly does not depend on the off-diagonal elements of the states' density operators. The error is reduced the closer the common latitude of states is to the equator of the Bloch sphere.

If the elements shown in figure \ref{fig} cannot form a POM (i.e. they are all in one half of the Bloch sphere), then the optimal strategy is the binary decision strategy \cite{Helstrom} which best discriminates between the pair of states with least overlap.

Should we add a fourth state to this picture, we run into some interesting problems. Those properties of $\hat{C}$ which were determined by (\ref{det}) for the original three states cannot change, as those conditions must still hold. If the fourth (\ref{det}) is not consistent with this, it can only lead to the situation where {\em no} $\hat{C}$ can satisfy both (\ref{det}) and (\ref{comp}) for all states, and some zero operator POM elements {\em must} be introduced. The three possibilities for adding a fourth state are:
\begin{enumerate}
\item The new state is on the common latitude of the three original states. Then the optimal $\hat{C}$ is the same as it was for three states.
\item The new state is in the opposite Bloch Sphere hemisphere from all three states. Then Yuen's solution \cite{Yuen} for states which themselves form a POM (when multiplied by suitable coefficients) applies and $\hat{C} \propto \hat{1}$.
\item The new state shares a hemisphere but not a latitude with all three original states. Then no solution exists which does not contain at least one zero operator POM element. The optimal strategy will have non-zero elements corresponding to the subset of three states whose common latitude is closest to the equator of the Bloch Sphere.
\end{enumerate}
Further additional states follow the same pattern. The subset of states which the optimal strategy detects will always either have common diagonal elements in some basis, or be able to form a POM when multiplied by suitable coefficients, or consist of only two states.

\section{Conclusions}

We can simplify the problem of finding the best strategy for discriminating between non-orthogonal quantum states by solving the necessary and sufficient conditions for minimum error (\ref{nscon}) for the error itself rather than solving directly for the optimal strategy. This is easier since then we have only one operator variable ($\hat{C}$) and there is no issue with variables not being independent.

When you do this is becomes obvious (we have $N$ equations for only one unknown operator) that, in general, some states will never be selected because the corresponding POM element will have to be a zero operator. The exceptions to this are when the number of states is small, or some limiting symmetry applies to all of them. It is interesting to note here that the subset of states which corresponded to non-zero POM elements in the example {\em had} to be one of the special symmetry cases. 

The example gives the form of the solution for the optimal discrimination between all possible sets of equally likely pure qubit states.
This example has also shown that the minimum error obtained in the symmetric states case \cite{Symerr} is more generally applicable. The corresponding solution described here applies to any arrangement of states with common diagonal elements, though this solution is {\em not} generally the square-root measurement as described in \cite{Symerr}. It can be shown that this result is still valid in higher dimensional bases.

We have also shown that the weights $\mathrm{Tr}(\hat{\Pi}_{k})$ of the POM elements are irrelevant to the optimality of the strategy. The only restriction on them is that the POM is complete (\ref{comp}) and thus realisable. This means that we never {\em need} more than $D^{2}$ ($D$ is the dimension of the system) possible outcomes and thus $D^{2}$ non-zero POM elements to achieve the minimum error. In any situation where we {\em could} use more than $D^{2}$ elements and still achieve the minimum error, the minimum error measurement strategy cannot be unique. There will then be an unlimited number of optimal strategies differing only in the weights of their elements. This constitutes a state discrimination equivalent of Davies theorem for the accessible information \cite{Davies}.

\begin{theacknowledgments}

This research was funded by EPSRC. I would also like to thank Professor Stephen M. Barnett for his help in introducing this field to me in the course of my PhD research.

\end{theacknowledgments}

\bibliographystyle{aipproc}

\bibliography{thesiscits}

\end{document}